\documentclass{article}

\usepackage{microtype}
\usepackage{graphicx}
\usepackage{subfigure}
\usepackage{booktabs} 

\usepackage[accepted]{icml2021}

\usepackage{hyperref}

\icmltitlerunning{The Influence of Explainable Artificial Intelligence}

\begin{document}

\twocolumn[
\icmltitle{The Influence of Explainable Artificial Intelligence:\\
Nudging Behaviour or Boosting Capability?}




\begin{icmlauthorlist}
\icmlauthor{Matija Franklin}{to}

\end{icmlauthorlist}

\icmlaffiliation{to}{Department of Experimental Psychology, University College London, London, UK}

\icmlcorrespondingauthor{Matija Franklin}{matija.franklin@ucl.ac.uk}

\icmlkeywords{Explainable Artificial Intelligence, XAI, Nudge, Boost, Influence, Behavioural Change}

\vskip 0.3in
]



\printAffiliationsAndNotice 

\begin{abstract}
This article aims to provide a theoretical account and corresponding paradigm for analysing how explainable artificial intelligence (XAI) influences people's behaviour and cognition. It uses insights from research on behaviour change. Two notable frameworks for thinking about behaviour change techniques are nudges - aimed at influencing behaviour - and boosts - aimed at fostering capability. It proposes that local and concept-based explanations are more adjacent to nudges, while global and counterfactual explanations are more adjacent to boosts. It outlines a method for measuring XAI influence and argues for the benefits of understanding it for optimal, safe and ethical human-AI collaboration.

\end{abstract}

\section{Introduction}

Deep Learning (DL) is a subfield of Machine Learning (ML) that focuses on developing Deep Neural Network (DNN) models \cite{shahroudnejad2021survey}. DNNs are complex models that are capable of achieving high performance on a variety of tasks. Many deep neural network models are uninterpretable black-boxes, which often leads to users having less trust in them \cite{miller2019but}. This lack of interpretability and trust can have negative consequences – people might use an AI that makes mistakes, or not use an AI that could improve the chances of a desired result. To improve the interpretability and trustworthiness of black-box models, Explainable Artificial Intelligence (XAI) research focuses on developing methods to explain the behaviour of these models in terms that are comprehensible to humans \cite{molnar2020interpretable}.

Explanations provided by XAI can impact human behaviour and cognition \cite{donadello2020explaining, dragoni2020explainable}. This paper aims to apply frameworks from research on behaviour change within the broader field of behavioural science \cite{ruggeri_behavioral_2018}. Two notable frameworks for thinking about and categorising behaviour change techniques are ‘nudges’ and ‘boosts’ \cite{grune2016nudge}. A nudge is any aspect of choice architecture – the context in which people make decisions - aimed at influencing the behaviours of individuals without limiting or forcing options \cite{sunstein2014nudging}. Boosts are interventions that foster people’s competencies through changes in knowledge and skills, so that they can make their own choices more effectively \cite{hertwig2017consider}. By doing so this paper seeks to put forward a framework and a paradigm for evaluating whether within the context of human-AI interaction an XAI is nudging performance or boosting capability. It will discuss the implications of each in relation to human-machine collaboration as well as the ethics of influence.

\section{Different explanations, different outcomes}

A great deal of XAI research has gone into developing methods to improve the explainability of DNN models \cite{molnar2020interpretable}. As it stands it is not fully clear how different methods may influence human behaviour.

Feature importance methods (such as saliency methods) generate scores that reveal how important a feature (like a word vector or pixel) is to the AI's decision-making process \cite{bhatt2020explainable}. Explanations generated by these methods can be either global or local in scope \cite{lundberg2020local}. Global explanations, like those provided by SHAP (SHapley Additive exPlanations) models, give a quantitative indication of the importance of each input variable on the model's output \cite{lubo2020machine}. Local explanation methods, such as LIME (Local Interpretable Model Agnostic Explanations), generate a numeric score showing the importance of an input variable in relation to the outcome variable \cite{lee2019developing}.

Counterfactual explanations show what the model's output would have been if one or more of the inputs had been different \cite{verma2020counterfactual}. This can be helpful in understanding why the model arrived at a particular output. Finally, concept-based explanations attempt to explain a model's output by referencing pre-defined or automatically generated sets of concepts that are comprehensible to humans \cite{kazhdan2020now}.

Even though there has been progress in the development of XAI models, it is still not completely clear which model should be used for human-machine collaboration, and for what purpose. There are still many open question. For example, do local explanations nudge performance in the short term but not necessarily provide enough information to educate a user and boost their capability in the long term? Do global explanations provide enough information to educate? Can counterfactual explanations teach people how their AI works or allow them to identify errors?

A systematic literature review of 241 papers looked at how the validity and usefulness of explanations have been evaluated by the authors of XAI methods \cite{anjomshoae2019explainable}. Most studies only conducted user studies in simple scenarios or completely lacked evaluations. The results show that 32\% of the research papers did not include any type of evaluation. Furthermore, 59\% of the research papers conducted a user study to evaluate the usefulness of the explanation (with a small minority also evaluating user trust towards the AI system). Finally, 9\% of the papers used an algorithmic evaluation, which did not involve any empirical user research. These initial findings suggest that different explanations will lead to variations in performance on a task \cite{lage2019evaluation, narayanan2018humans}, and will not always necessarily improve performance and understanding \cite{kindermans2019reliability}. If viewed as a behaviour change intervention, when does an XAI serve as a nudge, changing behaviour, and when does it serve as a boost, improving capability?

\section{Local and concept-based explanations as nudges}

 The common aim of nudges is to predictively change targeted behaviours. A nudge is any aspect of choice architecture aimed at influencing people’s behaviour, without limiting or forcing options, or significantly changing their economic incentives \cite{thaler2008nudge}. All environments influence behaviour to some extent, even when people are not aware of it. Intentionally changing choice architecture is nudging. Nudges take many shapes \cite{sunstein2014nudging}. Default rules such as automatic enrollment in programs, automate decision-making for people by setting a default. Simplification nudges reduce the amount of information presented to people to avoid information overload. The use of descriptive social norms - telling people what most other people are doing - influences behaviour. As a policy tool, nudging has been used in over 80 countries worldwide, and by major supranational institutions such as the World Bank and UN \cite{oecd2017behavioural}.

Nudges have been heavily influenced by Daniel Kahneman's dual-process account of reasoning \cite{kahneman2003perspective}. He proposed that people have "two systems" in their mind - System 1 and System 2. System 1 thinking is heuristic. It reacts intuitively and effortlessly, without analysing all available information. System 2 is an analytical and effortful, rationalising process. System 1 thinking is fast, and thus accounts for most behaviour. System 2 can re-evaluate System 1 thinking, thus using System 2 thinking leads to fewer erroneous decision. However, this is difficult, as it requires more cognitive effort. Importantly, some factors and contexts are more likely to trigger System 1 or System 2 thinking than others. Per \citet{sunstein2016ethics}, nudges work by targeting either System 1 thinking, thus influencing behaviour without the awareness of the decision maker, or System 2 thinking, thus promoting deliberative thinking. 

A famous example of nudging is using disclosure nudges. Disclosure nudges disclose decision-relevant information \cite{sunstein2014nudging}. They are educative, because they provide a learning experience, and target System 2, by promoting deliberative thinking. Disclosure nudges are rooted in three insights. First, as uncertainty promotes erroneous decision making \cite{kochenderfer2015decision}, disclosure nudges seek to reduce uncertainty with decision-relevant information. Second, when too-much decision-irrelevant information is present, people find decision-making more challenging \cite{rogers2013infobesity}; disclosing only decision-relevant information can, therefore, reduce error. Finally, disclosure nudges create an emphasis frame, making relevant information more salient \cite{chong2007framing}.

Viewed through this framework, local feature importance explanations and concept-based explanations can be viewed as a type of disclosure nudge. They will provide a small amount of decision-relevant information at the time a person needs to decide whether or not to trust an AIs advice or prediction. It thus could be the case that findings from the extensive body of research on disclosure nudges may generalise to local and concept-based explanations. Such comparisons can guide future practice and research directions. 

\section{Global and counterfactual explanations as boosts}

Boosts are interventions that aim to promote people’s competencies, so they can make better decisions \cite{grune2016nudge}. Proponents of boosts aim to foster skills and decision heuristics that can persist over time, throughout different decision contexts \cite{hertwig2017nudging}. An example of a boost is teaching people better decision-making skills with the use of decision trees \cite{hertwig2009fast}, or by teaching them how to calculate the expected value of a prospect \cite{franklin2019optimising}. Unlike nudges, a boosts effectiveness requires people’s awareness of the boost and motivation to improve their competencies. Furthermore, boosts differ from educational nudges, in that they do not guide people towards a decision, but rather they assume that better-skilled people will make advantageous decisions. In the context of a decision, boosts provide people with the right capability for the task at hand.

Boosts can either promote abilities that are specific to a single domain (e.g., decision-making under risk) or generalise across many domains (e.g., statistical literacy; \citet{hertwig2017nudging}). Furthermore, boosts can change the choice architecture to foster competencies, directly teach people skills, or do both. Boosts have been used to improve people’s decision-making capabilities by promoting people’s understanding of statistical information. This competence has been previously achieved through: using graphical representations of statistical information \cite{lusardi2017visual}, training math skills \cite{berkowitz2015math}, representing information that avoids framing effects \cite{gigerenzer2007helping}, and giving people simulation based representations of statistical information \cite{hogarth2015providing}.

Through the lens of behaviour change theories, global feature importance explanations and counterfactual explanations can be viewed as a type of boost. Counterfactual explanations can be used to train people to understand when their AI tool is more likely to err. This may generalise as a more broader knowledge about how AI tools "work" more broadly. Global explanations may increase people's knowledge about what factors are relevant for a given context. This can improve people's own predictions and competencies within a certain domain. As was the case with nudges, findings from the research literature on boosts may generalise to global and counterfactual explanations. These comparisons could guide both practice and future research on improving human-AI collaboration. 

\section{Evaluating XAI Influence}

In light of the previous discussion on certain XAIs as performance boosting and others as capability boosting, this section will propose  methods for testing whether a explanation is nudging performance or boosting capability. It will outline human-AI interaction paradigms for evaluating an XAI's influence. 

Measures of performance will be highly context-specific \cite{hoffman2018metrics}. It is also possible to identify whether explanations improve performance by helping people do better, or by reducing the number of mistakes people make \cite{franklin2022human}. Ideally, performance on the same task without an AI will be compared to performance on a task with an AI, and with an XAI. Multiple different XAIs can be used in this comparison. It could either between-subject - comparing the performance of people across different groups - or within-subject - tracking the performance of an individual across time.

To give an example, one could compare the effects of local and global explanations for improving a doctor's diagnostic ability. Specifically, the study would evaluate the benefits of an AI diagnostic tool on performance, and see how this improves when people receive explanations for how that diagnostic tool makes decisions. In this example, doctors will be tested on their ability to detect pneumonia from x-ray images. In a between-subject paradigm, doctors could be randomly placed in four separate groups: 1) doctors with no AI tools, 2) doctors with AI tools, 3) doctors with AI tools and a local explanation, 4) doctors with AI tools and a global explanation. The difference between the first and second group would be a pure measure of how the AI tool improves pneumonia detection. The difference between the second and third, or the second and fourth group would allow one to look at the benefits of the explanation on performance. Finally, the difference between the third and fourth group would serve as a measure of what XAI is more effective in improving performance. 

A similar paradigm can be used to measure capability. An AI without an explanation can allow people to understand the AI's decision-making process by detecting patterns \cite{peltola2019machine}. XAI can provide a more direct learning experience for users. This learning could be \textit{procedural}, for example a change in a certain ability, or \textit{semantic}, for example an increase in factual knowledge \cite{berliner2013handbook}. Procedural knowledge can be measured as people's performance in the absence of an AI tool after using an AI tool. The pause between initially using the AI tool and then measuring performance without it can be more or less longitudinal. Longer time frames would make for a more valid measure as they would reduce the confounding impact memory may have on performance. Back to the previous example, each doctor can be brought back for testing 2 weeks after the initial study. In this case they would be performing the same activity - pneumonia detection from x-ray images - but this time with no AI tools or explanations. A lack of differences between their previous and current performance would suggest that their interaction with the AI or XAI models led to an increase in capability. Measuring semantic knowledge with a similar paradigm  would involve direct, context-specific questions to test a doctor's knowledge.

Finally, measuring changes in people's mental models can provide useful information for whether an XAI has led to new knowledge and capability. Mental models are representations of a person's understanding of some system or object \cite{lagnado2021explaining}. An XAI could change the person's mental models about the task, domain or the AI tool. Given that people are able to infer causal structures from explanations \cite{kirfel2021inference}, explanations can both establish the presence and change the direction of perceived cause and effect relationships. Mental models can be measured using a nearest neighbor task, where participants select the explanation or diagram that best fits their beliefs, or with concept mapping, where users create a diagram which outlines their knowledge \cite{hoffman2018metrics}.

\section{Issues with XAI influence}

Not understanding how an XAI's explanation influences behaviour and cognition raises three broader issues. First, there may be individual differences in the way explanations influence specific users on certain tasks \cite{tomsett2018interpretable}. A review of 137 articles on XAI applied to different domains shows that although there is preliminary evidence for visual explanations being more acceptable to everyday users, most studies have been directed at expert users \cite{islam2022systematic}. More research evaluating everyday users' reactions to XAI is needed \cite{franklin2022human}. The person receiving the information matters.

Second, it is not evident what amount of and type of information a user should receive. Detailed explanations might conceal more information than they reveal. This phenomenon is known as \textit{information overload} in psychology – when a person receives too much information, or more specifically when the amount of input to a system exceeds its processing capacity \cite{sutcliffe2009information}. This is also true even when all of the information is directly relevant to the task at hand. Recent research has found that information overload effects can occur in response to explanations provided by XAI methods \cite{ferguson2022explanations}. In general, it was found that the more detailed the explanation, the less useful and trustworthy it was considered to be. Specifically, non-numerical explanations in plain English that listed the variables involved in a model were rated as more understandable and trust-worthy than SHAP models. 

Other research has found the opposite to be true for highly technical individuals. A 2021 study found that people with backgrounds in AI have different requirements for explanations \cite{ehsan2021explainable}. Specifically, individuals with a background in AI get more use out of numerical explanations compared to less technical individuals. This means that it is not possible to make generalizations, such as that more information will always lead to better decisions. 

Finally, persuasive explanations will influence behaviour and preferences in ways that are currently difficult to predict. Current research suggests that an XAI tool, as opposed to a regular AI tool, is more likely to produce a change in behaviour \cite{donadello2020explaining, dragoni2020explainable}. Researchers have also developed XAI methods which can generate misleading explanations that can both increase trust in the AI and mislead domain experts \cite{lakkaraju2020fool}. It is therefore essential to understand the direction of the behaviour change, whether it is desirable, and how this will differ from one XAI method to another. This is especially relevant given the fact that preference and behaviour have a bidirectional causal relationship \cite{ariely2008actions}, which means that AI systems can influence preferences by changing behaviours \cite{ashton2022problem, franklin2022recognising}. By pushing behaviour in a certain direction, over time, people's preferences for what to do and how to do it can change.

\section{Conclusions}

This paper proposed a framework for understanding XAI influence, and a method for measuring it. It proposed that local feature importance explanations and concept-based explanations are adjacent to disclosure nudges, and global feature importance explanations and counterfactual explanations are adjacent to boosts. It is thus possible that the extensive research conducted on nudges and boosts can provide insights for how XAI will influence behaviour. Translational research using existing insights from AI, Behavioural Science and Human-Computer Interaction  would be beneficial. Such research could expand our understanding of influence dynamics in human-AI interaction.

There are at least three reasons for why understanding XAI influence is important for better human-AI collaboration and teaming. First, the adequate use of explanations allows people to identify AI errors; namely due to spurious correlations \cite{sagawa2020investigation} or runaway feedback loops \cite{ensign2018runaway}. This is especially relevant in light of recent work on making models robust to spurious correlations by leveraging humans’ common sense knowledge of causality \cite{srivastava2020robustness}. Identifying XAIs for promoting this error-detecting ability would be beneficial.

Second, it is important to understand which XAIs result in unwanted or negative (i.e., capability and performance decreasing) influence or \textit{sludge} \cite{thaler2018nudge}. Sludge can encourage self-defeating behaviours or discourage a person's best interest. A common form of sludge is friction by making something slower or unnecessarily complicated \cite{sunstein2018sludge}. Research showing that certain forms of XAI result in information overload is an example of this \cite{ferguson2022explanations}. \citet{sunstein2020sludge} argues for sludge audits whereby one identifies the sources of sludge and eliminates them. Similar practices should be used in XAI. 

Finally, understanding the influence of XAI can allow for a more optimal distribution and selection of XAI methods. Some people performing certain tasks will want performance boosts. Others will want to learn. Understanding XAI influence would allow us to match these preferences.

\bibliography{main.bib}
\bibliographystyle{icml2021}

\end{document}